# Light extraction from 2D materials using liquid formed micro-lenses

C S Woodhead[1], J Roberts[1], Y J Noori[1], Y Cao[1], R Bernardo-Gavito[1], P Tovee[1], A Kozikov[2], K Novoselov[2] and R J Young[1]

[1] Physics Department, Lancaster University, Lancaster, Lancashire, LA1 4YB
[2] School of Physics and Astronomy, University of Manchester, Oxford Road, Manchester, M13 9PL

E-mail: r.j.young@lancaster.ac.uk

**The recent discovery of semiconducting two-dimensional materials has led to the prediction of a revolution in the field of optoelectronics, driven by the introduction of a series of new components that are just a few atoms thick. Key remaining challenges for producing practical devices from these materials lie in improving the coupling of light into and out of single atomic layers, and in making these layers robust to the influence of their surrounding environment. We present a solution to tackle both of these problems simultaneously, by deterministically placing a micro-lens directly onto the surface of these materials. These lenses are dynamically tuned to increase the coupling of light, whilst controlling chromatic aberration, before being set in place with UV light. We show that this approach enhances photoluminescence of tungsten diselenide ($WSe_2$) monolayers by up to 300%, and nearly doubles the imaging resolution of the system. Furthermore, this solution fully encapsulates the monolayer, preventing it from physical damage and degradation in air. The optical solution we have developed could become a key enabling technology for the mass production of ultra-thin optical devices, such as quantum light emitting diodes.**

## INTRODUCTION

Recent work into two-dimensional materials[1] has shown that their fundamental properties can be exploited to create novel optoelectronic devices[2-6], such as ultra-thin light sources[3]. Defects inside these materials have been shown to be capable of emitting single photons[2,7]. These advances are paving the way to produce electrically driven single photon sources that can be integrated into silicon, overcoming multiple technological barriers for photonic quantum information processing.  However, an outstanding challenge with devices based on these materials lies in maximising the photon extraction efficiency. Devices such as the two-dimensional LED pioneered by Withers et al. in [3], use transition metal dichalcogenides (TMDCs) to emit light, however this leads to a low extraction efficiency due to omnidirectional emission and gives poor quantum yields from the TMDC[8]. Recent studies have tried to improve the photoluminescence (PL) of these materials by either using chemical treatment via organic superacids[8], or incorporating plasmonic structures, such as gold nano-spheres[9], and gold coated trenches[10]. The gains from these methods, however, predominantly come from increases in the absorption and internal quantum efficiency coupled with a decrease in the radiative lifetime; therefore despite producing more light they do not directly address the problem of low extraction efficiency.

In solid state light sources, the efficiency of extraction can be improved using a simple technology, solid immersion lenses (SILs)[11-16]. SILs are typically formed from glass or plastic and are mounted directly onto the surface of a semiconductor device to improve optical properties, and optimise the extraction efficiency. Whilst these have proven to be effective solutions in conventional optoelectronic applications, they are difficult to implement in conjunction with surface emitters, such as devices based on TMDC monolayers. This is because the process of mounting them can cause damage to the monolayers, whilst unintentional air gaps at the interface can severely degrade performance. In this work we investigate the application of a recently developed form of SIL which uses a UV-curable epoxy[11], and directly employ it to TMDC materials. We show that such SILs can enhance the extraction efficiency from some of these materials and protect them against aging when subject to ambient conditions.

The fabrication process of epoxy SILs was demonstrated by Born et al. in [11], who showed that SILs could be deterministically positioned and formed by dispensing liquid UV-curable epoxy onto a substrate. The resultant optical element is shaped using electro-wetting[17] to achieve the required geometry and subsequently solidified by exposure to UV light. The dispensing environment is filled with a liquid phase medium, designed to increase the contact angle between the droplet edge and the substrate by modification of the interfacial surface tensions[18]. This form of SIL has been shown to enhance the performance of optical wireless imaging receivers based on classical semiconductor devices[19], but their application to monolayer materials has yet to be demonstrated, and interactions between these material systems explored.

Two different geometries of SIL have been predominantly studied; those with a hemispherical shape (h-SILs) and those with a Weierstraß shape (s-SILs)[20]. The latter form has a higher magnification than h-SILs, scaling as the refractive index of the SIL squared ($n^2$) as opposed to a direct linear relationship for a h-SIL ($n$). s-SILs have the advantage of increasing the light input/output coupling efficiency by being able to refract rays at the SIL-air boundary, thus collecting/delivering more light to and from a device. The main disadvantage of an s-SIL is its chromatic aberration, as the increased height leads to more variation in the path length taken by different rays. This can be very problematic for good quality imaging of a device, but acceptable for emitters such as quantum LEDs, where maximising light extraction is absolutely essential. By forming a SIL out of liquid epoxy, the geometry can be diversely modified to suit the application, by simply changing the filler solution[18], or using electrowetting[19], making it an extremely versatile method. This enables SILs to be produced which can provide the optimal trade-off between enhancement and chromatic aberration for the selected application. The output coupling efficiency of a SIL has been described by Moehl et al. in [13], as a product of the reduction in reflection losses ($K_T$) due to a more closely matched refractive index, and an increase in the solid angle in which photons can be collected ($K_\theta$). For the case of putting a SIL on a monolayer, $K_T$ can be ignored as the TMDC thickness is much less than the wavelength of emission, resulting in negligible reflection losses. The solid angle increase that occurs due to a SIL derives from the refraction that occurs at the SIL-air interface, for this reason h-SILs will not increase light output for a surface emitter due to negligible refraction. Any SIL with a height between that of an s-SIL and an h-SIL will have an enhancement at this interface. To calculate the enhancement due to $K_\theta$, the angle of refraction at the SIL-air interface ($\gamma$) needs to be calculated as a function of the release angle of the light from the monolayer ($\theta$). This is shown below in equation 1, where r is the radius of the SIL and a is the minimum distance between the base of the SIL and its spherical centre (h-r), $n_1$ is the refractive index of the SIL and $n_2$ is the refractive index of the environment (unity for air),

$$\gamma = \sin^{-1}\left[\frac{a}{r}\sin(\theta)\right] - \sin^{-1}\left[\frac{n_1 a}{n_2 r}\sin(\theta)\right] + \theta \tag{1}$$

This equation can be used to find the maximum angle of emission that can be coupled into the collection optics. A ray trace simulation is shown in figure 1, highlighting the increase in the number of rays that can be launched into the collection optics with a SIL present on the surface.

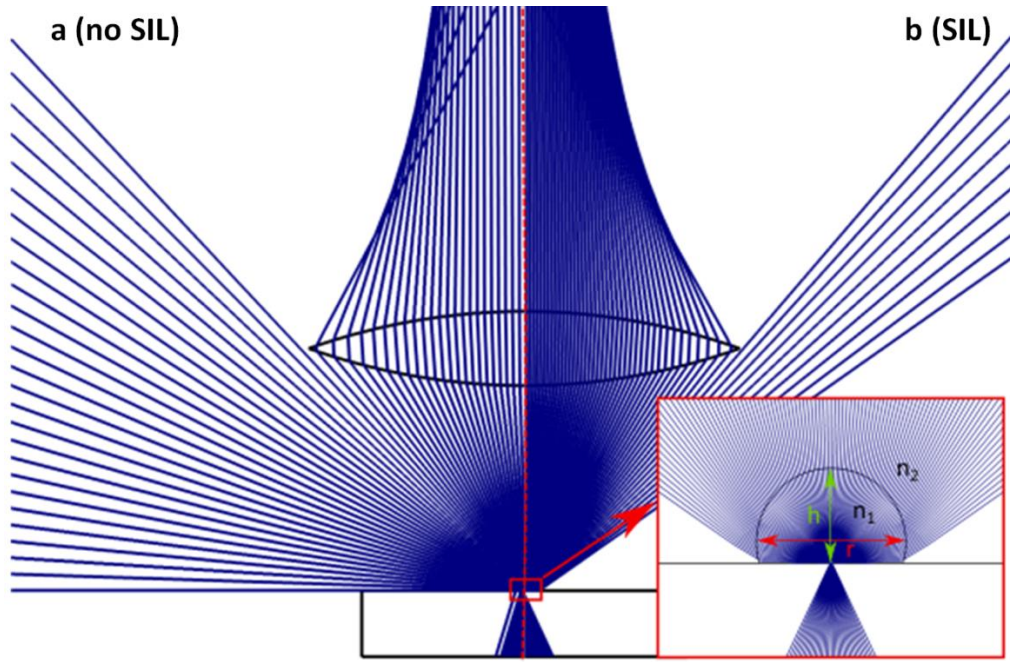

**Figure 1** A simulation illustrating how rays from a surface emitter couple to an optical system, with **(a)** and without **(b)** a solid immersion lens (SIL) in place. The SIL increases the coupling efficiency of rays emerging from the emitter. The inset shows a magnified region surrounding the SIL.

**MATERIALS AND METHODS**

In this study the optical properties of monolayers of both molybdenum disulphide ($MoS_2$) and tungsten diselenide ($WSe_2$) on the surfaces of $Si/SiO_2$ wafers were studied both before and after the application of an epoxy SIL. Monolayers were mechanically exfoliated[21-23] from natural/synthetic bulk crystals of $MoS_2/WSe_2$ respectively, and transferred to the surface of silicon wafers with a 290 ± 5 nm $SiO_2$ coating (the process is illustrated in Figure 2a).

To analyse the optical properties of individual monolayers, the samples were placed in an evacuated cryostat and excited with a 532nm CW laser through a 50x IR microscope objective lens with an NA of 0.65. Photoluminescence was collected from the sample at 300 K and 20 K through the same objective lens and passed through a single-stage spectrometer to a silicon EMCCD detector for wavelength-resolved measurements. The microscope objective was mounted on a piezo-driven XYZ stage with 20 nm step resolution, to produce photoluminescence maps as a function of position on individual monolayers.

Following the identification and optical analysis of appropriate monolayers, the epoxy SILs were produced and aligned to the monolayers using the following procedure, as illustrated in Figure 2b-d. The samples were first placed into a bath of glycerol, which acts as the filler solution that enables high contact angle droplets to form on the sample. The UV-curable epoxy (Norland adhesive 81) was dispensed onto the wafer through a 32 gauge needle, using an air dispensing system to accurately control the droplet size (Figure 2c). The needle was positioned above previously studied monolayers using a custom built stage with an endoscope camera, allowing relocation accuracy at the micron scale. A bias between the needle and sample could then be applied to tune the contact angle of the SIL, if required[11]. Finally, once the size and shape of the SIL was optimised, the needle was retracted and the epoxy was cured using an external UV light source (Figure 2d). Following the successful mounting of SILs onto previously studied monolayers, the samples were then re-measured using the same optical procedure.

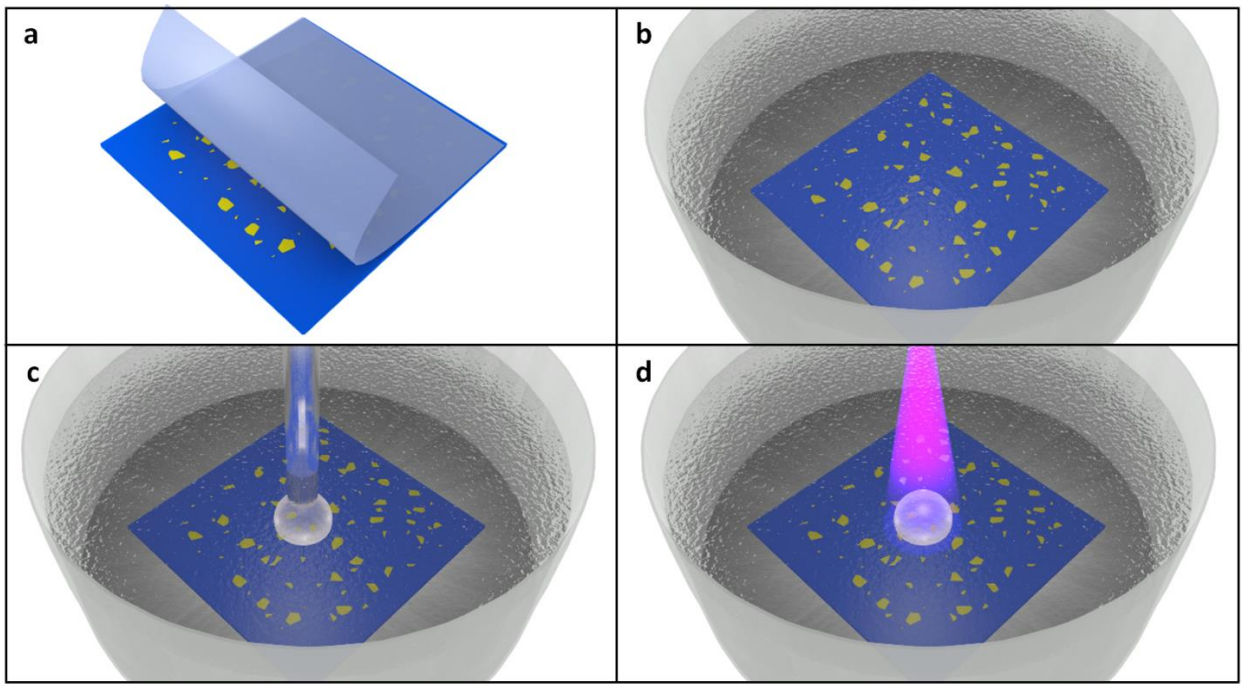

**Figure 2** The experimental process used to mount SILs deterministically on to monolayers of 2D materials; **(a)** monolayers were produced via mechanical exfoliation and then transferred onto $SiO_2$; **(b)** sample was immersed in a bath of glycerol; **(c)** UV curable epoxy was dispensed over the monolayer and shaped; **(d)** the epoxy was bonded onto the surface with a UV light source.

**RESULTS AND DISCUSSION**

Figure 3 shows optical microscope images taken of a flake of $WSe_2$, with both monolayer and bilayer thick regions, before and after mounting a SIL on top of it. The mounted epoxy SIL had a radius of 550 ± 5µm, and a height from apex to base of 700 ± 5 µm. The optical magnification through the SIL was measured to be 1.80, this helps to resolve fine features of the flake, as shown by the cracks in the flake which are now clearly visible. Optical artefacts visible in Figure 3d are a result of using a narrower aperture in the microscope to take the image. It can also be observed that a few small bubbles have become encapsulated, but providing the bubbles are not directly over the monolayer, these defects will have no effect on the light collection. Additional SILs with very similar dimensions, were mounted onto flakes of $WSe_2$ and $MoS_2$, giving similar enhancements in magnification.

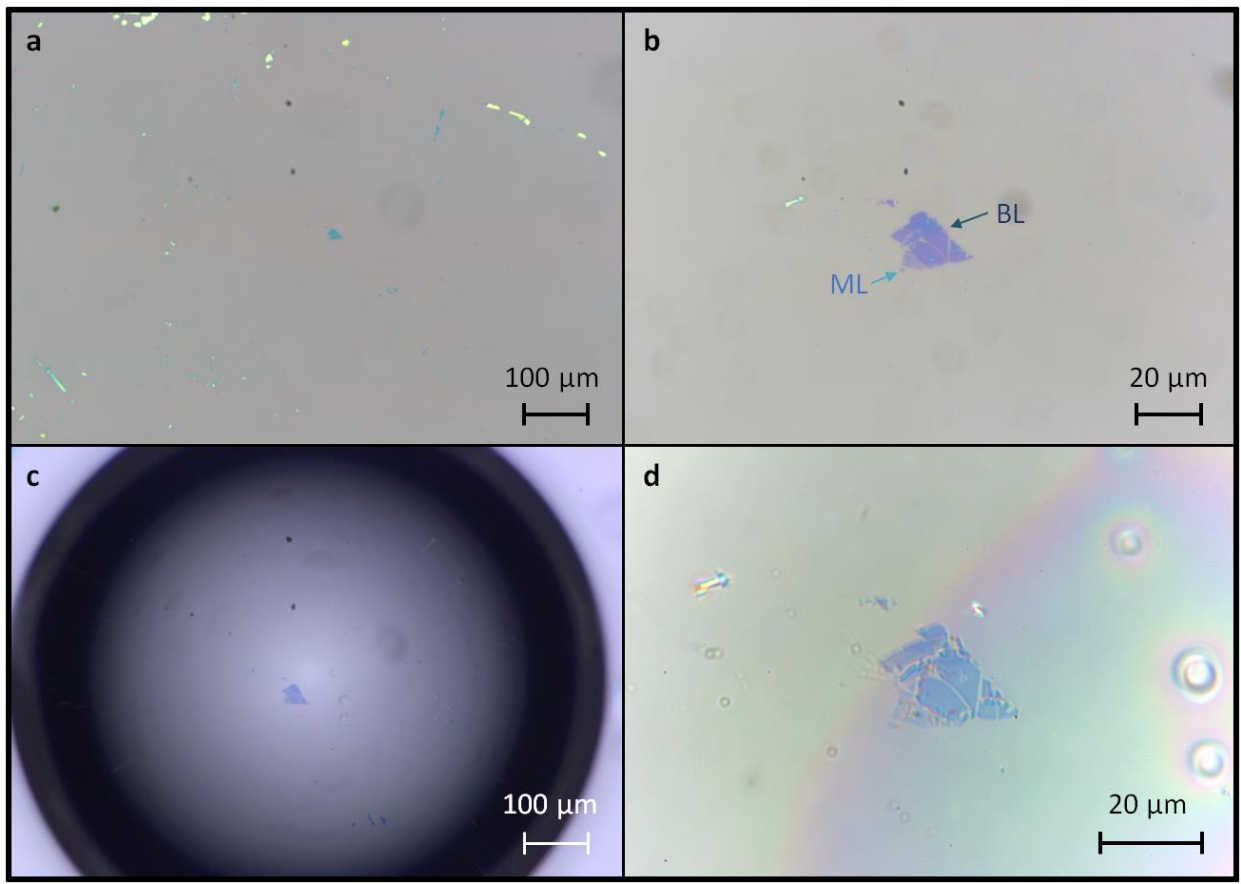

**Figure 3** Optical microscope image showing an isolated monolayer of $WSe_2$, before **(a,b)** and after **(c,d)** mounting a SIL; the scale bar in (d) has been adjusted to account for the SIL's magnification.

To assess the performance of the SIL, photoluminescence intensity maps of the $WSe_2$ monolayer (from Figure 3) were taken both before and after mounting a SIL, the results are shown in Figure 4. It is clear from Figure 4 that there is a significant increase in enhancement of the PL between (a)-(b), due to the mounting of the SIL. The extra resolution this provides enables more detail to be captured in PL maps, making edges and intensity differences easier to see. It should be noted that the horizontal artefacts seen are due to the PL maps being interpolated, to enable a better comparison with the optical images. The increase in resolution seen not only arises due to the magnification of the flake but also from a slight reduction of the laser's spot size. Theoretically the increase in resolution from a h-SIL in µPL arises from a reduction in laser spot size, which is proportional to $1/n$ (where n is the refractive index of the SIL)[24]. The laser acting as a plane wave will approximately project an airy pattern with a half-width at half-maximum (HWHM) of:

$$\text{HWHM} = \frac{0.26\lambda}{n\text{NA}_{obj}} \quad (2)$$

where n = refractive index of the volume above the TMDC, $\text{NA}_{obj}$ = numerical aperture of the µPL system and λ = excitation wavelength[13]. Calculating (2) with the experimental parameters here gives an expected resolution increase of 1.6 times. The value measured from Figure 4 is slightly larger than this, at 1.8x, this discrepancy is attributed to the SIL having a contact angle to the substrate greater than 90°. The overall improvement in magnification is thus somewhere between that of an h-SIL (linear dependence of n) and a s-SIL (quadratic dependence on n)[24]. This result shows that SILs in between the h and s-SIL geometries can give optical properties that are a combination of the two, with the studied SIL showing a greater magnification than an h-SIL without introducing strong chromatic aberrations. This suggests that gradually increasing the contact angle to the substrate from 90° will change the magnifications dependence on refractive index from a reciprocal to an inverse square relationship. This means that epoxy SILs have the potential to be formed into geometries that can provide the most optimised trade-offs between resolution and image quality.

Figure 4c shows a map taken at cryogenic temperatures (20 K); it can be seen that the temperature change has had no detrimental effect on the optical properties of the SIL, or flake. This result shows that the SILs are highly resistant to temperature changes, and can thus be used for photoluminescence studies at cryogenic temperatures. Some example spectra from (c) are shown in Figure 4d, these correspond to areas where there was some overlap between monolayer and bilayer, as observed by the two main peaks. Sharp exciton-like lines can be observed in each of these spectra, this is unusual as these sharp peaks do not appear anywhere else on this flake, nor do they appear at room temperature. Since these peaks all occur at the same wavelengths, it is unlikely that they are due to quantum dots, but could be due defects/impurities that are more likely to occur at the monolayer-bilayer boundary.

When comparing Figure 4a to Figure 4b there is a significant increase in the light collected across the sample, with the integrated intensity increasing on average by 4.0 times for the same excitation power. Figure 5a shows one of the PL spectra taken both before and after the application of the SIL, the fourfold enhancement in the integrated intensity is easily observed as a large increase in the peak. It was noticed that the relative enhancement varied with excitation power, this is shown in Figure 5b where the ratio between SIL to no SIL is plotted vs excitation power for different monolayer positions. The PL enhancement measured was twice as high as predictions from classical optics, using equation (1) and extrapolating into solid angles. This increased enhancement could possibly arise from constructive interference or confinement effects happening at a microscopic level, which cannot couple into the collection optics without a SIL. Simulating and quantifying these interactions are challenging due to the monolayer and the silicon dioxide substrate having dimensions that place the system into a semi-classical regime, making it computationally difficult to simulate. However, despite not fully being able to model the small scale light interactions, other emission increasing mechanisms such as doping of the sheet due to the epoxy can be discounted, as similar $WSe_2$ monolayers show no enhancement in their PL when a film of epoxy is cured above them (Figure ).

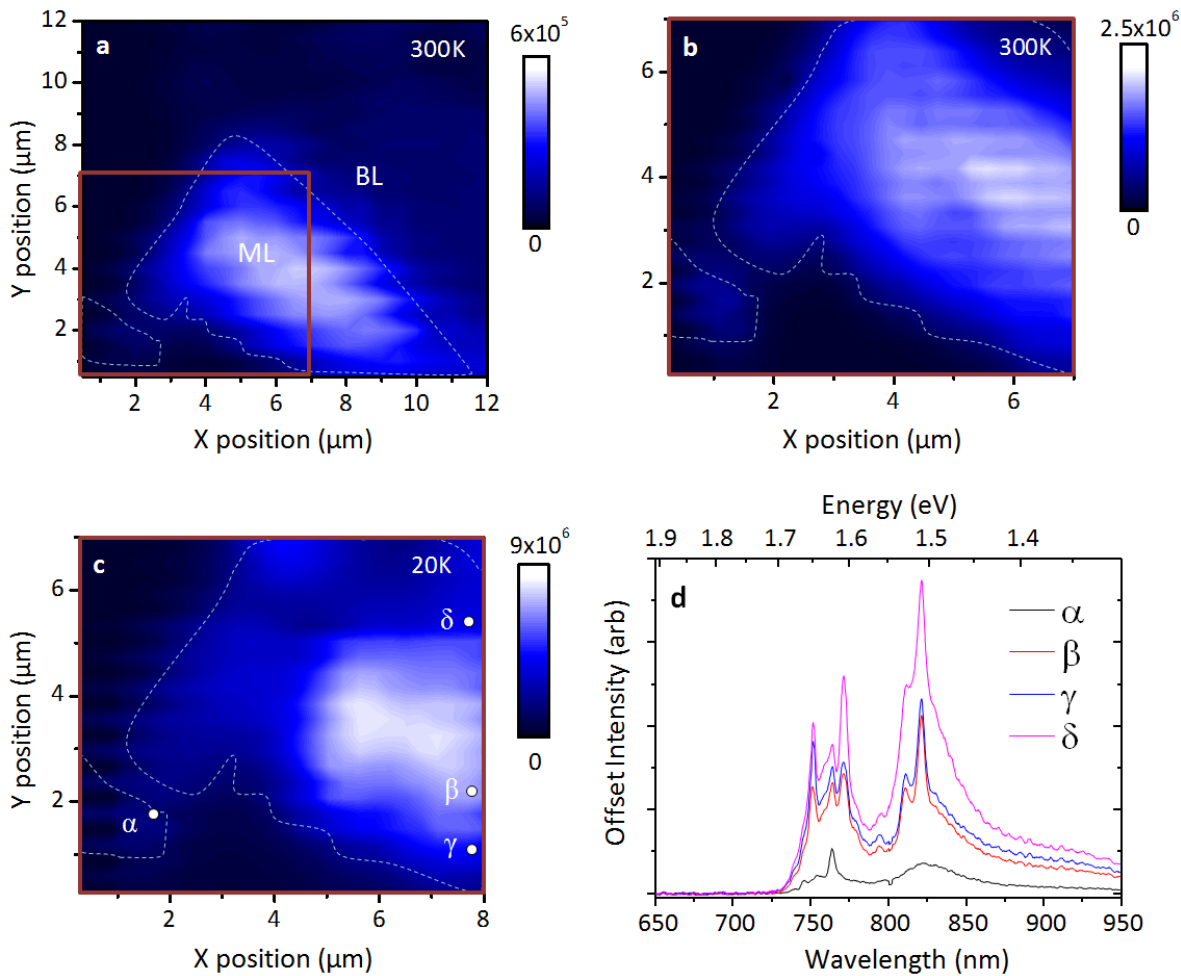

**Figure 4** Photoluminescence (PL) maps of a $WSe_2$ flake (outlined) taken both with and without a SIL and at low (20K) and high (300K) temperatures, as labelled; panels **(a)-(c)** show PL maps where the colour scale refers to the integrated intensity as a function of position; **(d)** shows example spectra from positions labelled in **(c)**. The orange box in **(a)** highlights the area seen in **(b)** and **(c)**.

The decrease of the intensity ratio with power is likely explained by the reduction in PL laser spot size with the presence of the SIL, and subsequent increase in power density seen by the monolayer. The increase in power density from the smaller spot size can increase PL intensity through charge screening, leading to a reduction of relative intensity with power. A decreased spot size could also cause an increase in localised heating of both the epoxy and the flake, which may affect both the emission of the monolayer and the optical properties of the SIL due to thermal expansion.

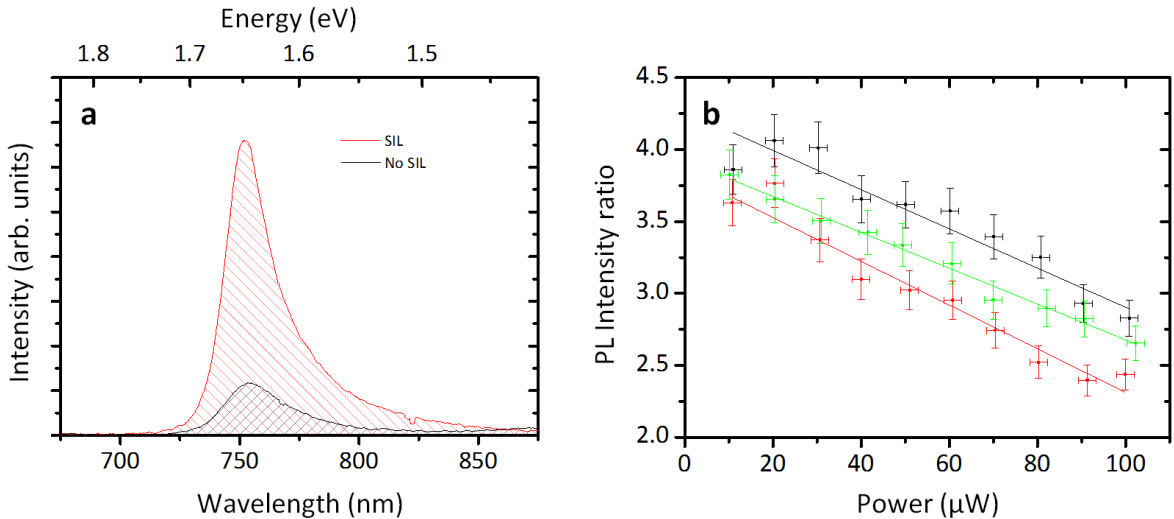

**Figure 5** Graphs showing **(a)** example PL spectra for monolayer $WSe_2$ at 300K excited using a 10µW laser both with and without a solid immersion lens (SIL); **(b)** the power dependence of the ratio in PL intensity of SIL to no SIL for three different positions on the monolayer.

SILs were also mounted on top of $MoS_2$ monolayers to see if they could be used to enhance the optical properties of the device in a similar fashion as for $WSe_2$. Upon mounting a SIL the emission of the monolayers was found to be heavily suppressed, and to investigate why, a 1.5 µm thick layer of epoxy was spin-coated onto the samples containing both types of monolayer. PL spectra from these samples are compared in Figure , where it can be observed that there is a significant reduction in PL intensity for the epoxy on $MoS_2$, with an accompanied red shift in the central wavelength. On the $WSe_2$ no significant change in the peak intensity was observed, however central wavelength blue-shifts, which is accompanied by a narrowing of the peak. The shift of the $WSe_2$ peak and its sharpening are possibly attributed to a change in the proportion of emission from charged excitons, causing a reduction in the monolayer peak intensity at longer wavelengths. This would indicate that the monolayer $WSe_2$ can be doped by the epoxy. These effects of doping also allow us to explain the results observed for $MoS_2$. The mechanically exfoliated $MoS_2$ is known to be naturally n-type[25], and it has been previously stated by Li et al. in [26] that additional n-type doping results in a quenching of the PL. The epoxy bonded to the flake may be introducing surface dopants resulting in the observed reduction in PL. $MoS_2$ monolayers are highly sensitive to doping, with similar effects being reported to occur between $MoS_2$ monolayers and the substrate that they are placed on[27].

The UV curing process of the epoxy may also induce compressive strain, influencing the optical properties of TMDCs. It has been shown that compressive strain can change the bandgap of $MoS_2$ from direct to indirect[28], which would decrease the emission. The compressive strain values needed to induce a shift from direct to indirect bandgaps for $MoS_2$ and $WSe_2$ are 0.5% and 1.5% respectively[29]. Since the threshold for WSe$_2$ is higher than MoS$_2$, the epoxy may have induced enough compressive strain to change the bandgap in MoS$_2$, but not in WSe$_2$.

It was hypothesised that in addition to the improved optical properties, the SIL may also increase the longevity of TMDC monolayers in ambient conditions. Gao et al. in [30] showed that under ambient conditions monolayer TMDCs degrade due to oxidation and the introduction of organic contaminates, but they themselves demonstrated encapsulation in transparent polymers can help prevent this. In our experiment over a time period of 2 months, there was no observable change in any of the encapsulated monolayers, despite being left in ambient conditions. Furthermore, the PL spectra remained identical for the same excitation powers, showing that no degradation occurred during this time period.

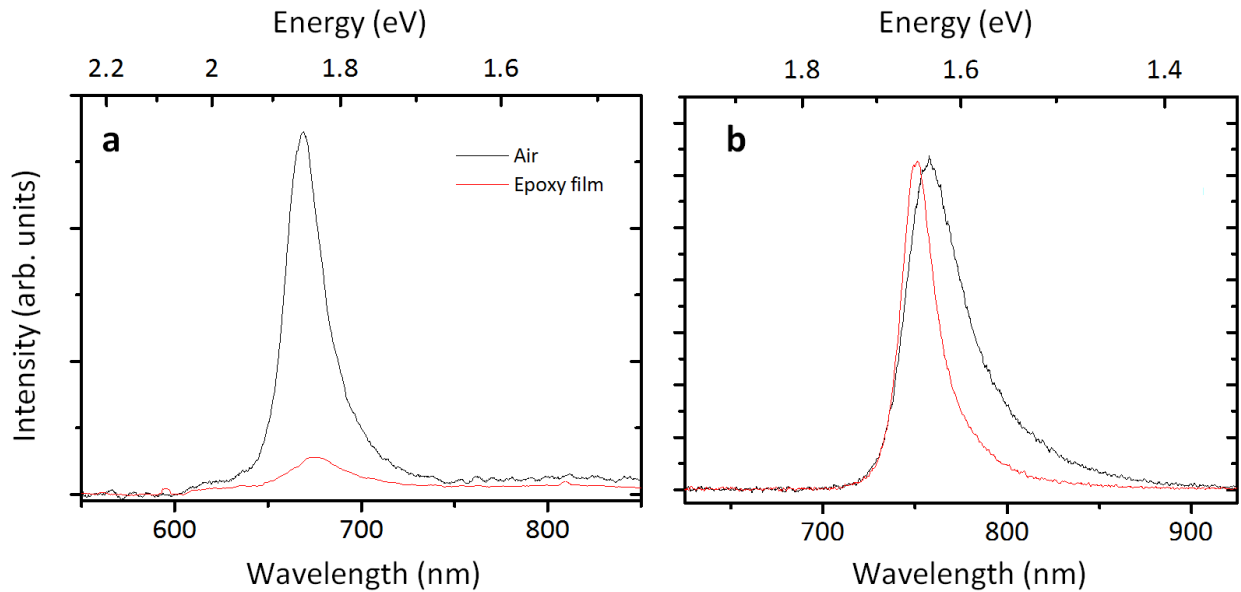

**Figure 6** PL spectra of **(a)** MoS2; and **(b)** WSe2 before and after the application of a thin film of cured epoxy.

**CONCLUSIONS**

In conclusion, we have shown that UV-curable epoxy based solid immersion lenses can be used to greatly increase the photoluminescence of WSe$_2$ monolayer sheets, with the measured photoluminescence intensity increasing by 3.5-4.0 times at low excitation powers. This was higher than what was theoretically expected, and though several explanations were explored, further investigation into this effect is needed to confirm how this enhancement arises. The magnification provided by the SIL was found to be approximately 1.8 times, enabling photoluminescence maps to be produced with a higher resolution. This enhanced magnification and resolution could be of excellent use in the optical analysis of WSe$_2$, allowing for more details to be observed, and collection optics to be simplified. Conversely, MoS$_2$ showed a quenching in the PL when the epoxy was incorporated. It is likely that the epoxy is either chemically altering the sheet or introducing compressive strain, however the mechanism behind this is unclear. Depending on whether the epoxy is attacking the transition metal, chalcogen or introducing a strain field, will determine what TMDC materials can be enhanced using epoxy based SILs.

Epoxy SILs are an excellent enabling technology for $WSe_2$ based optoelectronic devices such as LEDs. The SILs should increase the optical coupling efficacy of these structures by at least 100%, but with a perfect s-SIL potentially up to 400%. The great advantage of this increase in light extraction is that it will interact in an accumulative manner with any other emission increasing methods, e.g. photonic crystal cavities, or mirrors. This makes the epoxy SILs an important component for achieving close to unity input/output coupling efficiency for this type of device. Furthermore, SILs can help protect monolayer devices from degradation in air and the epoxy's insulating properties mean that it should have very little influence on the electrical properties of the LED. Overall, epoxy SILs are an incredibly promising optical package for the future of $WSe_2$ based optoelectronics.

**ACKNOWLEDGEMENTS**

RJY acknowledges support by the Royal Society through a University Research Fellowship (UF110555). This work was also supported by grants from The Engineering and Physical Sciences Research Council in the UK.

1. Castellanos-Gomez A. Why all the fuss about semiconductors? *Nature Photon* 2016; **10:** 202 - 204.

2. Berraquero CP, Barbone M, Kara DM, Chen X, Goykhman I, Yoon D*, et al.* Atomically thin quantum light emitting diodes. 2016**:** 160308795P.

3. F.Withers, Pozo-Zamudio OD, Mishchenko A, Rooney AP, Gholinia A, K.Watanabe*, et al.* Light-emitting diodes by band-structure engineering in van derWaals heterostructures. *Nature Mater* 2015; **14:** 301 - 306.

4. Sun Z, Martinez A, Wang F. Optical modulators with 2D layered materials. *Nature Photon* 2016; **10**.

5. Pospischil A, Humer M, Furchi MM, Bachmann D, Guider R, Fromherz T*, et al.* CMOS-compatible graphene photodetector covering all optical communication bands. *Nature Photon* 2013; **7:** 892 - 896.

6. Wang QH, Kalantar-Zadeh K, Kis A, Coleman JN, Strano MS. Electronics and optoelectronics of two-dimensional transition metal dichalcogenides. *Nat Nano* 2012 11//print; **7**(11)**:** 699-712.

7. KoperskiM, NogajewskiK, AroraA, CherkezV, MalletP, Veuillen JY*, et al.* Single photon emitters in exfoliated WSe2 structures. *Nat Nano* 2015 06//print; **10**(6)**:** 503-506.

8. Amani M, Lien D-H, Kiriya D, Xiao J, Azcatl A, Noh J*, et al.* Near-unity photoluminescence quantum yield in MoS2. *Science* 2015; **350**(6264)**:** 1064.

9. Sobhani A, Lauchner A, Najmaei S, Ayala-Orozco C, Wen F, Lou J*, et al.* Enhancing the photocurrent and photoluminescence of single crystal monolayer MoS2 with resonant plasmonic nanoshells. *Appl Phys Lett* 2014; **104**(3)**:** 031112.

10. Wang Z, Dong Z, Gu Y, Chang Y-H, Zhang L, Li L-J, *et al.* Giant photoluminescence enhancement in tungsten-diselenide-gold plasmonic hybrid structures. *Nat Commun* 2016 05/06/online; **7**.

11. Born B, Landry EL, Holzman JF. Electrodispensing of Microspheroids for Lateral Refractive and Reflective Photonic Elements. *IEEE Photonics Journal* 2010; **2**(6)**:** 873-883.

12. Liu Z, Goldberg BB, Ippolito SB, Vamivakas AN, Ünlü MS, Mirin R. High resolution, high collection efficiency in numerical aperture increasing lens microscopy of individual quantum dots. *Appl Phys Lett* 2005; **87**(7)**:** 071905.

13. Moehl S, Zhao H, Don BD, Wachter S, Kalt H. Solid immersion lens-enhanced nano-photoluminescence: Principle and applications. *J Appl Phys* 2003; **93**(10)**:** 6265-6272.

14. Yoshita M, Sasaki T, Baba M, Akiyama H. Application of solid immersion lens to high-spatial resolution photoluminescence imaging of GaAs quantum wells at low temperatures. *Appl Phys Lett* 1998; **73**(5)**:** 635-637.

15. Young MP, Woodhead CS, Roberts J, Noori YJ, Noble MT, Krier A, *et al.* Photoluminescence studies of individual and few GaSb/GaAs quantum rings. *AIP Advances* 2014; **4**(11)**:** 117127.

16. Zwiller V, Björk G. Improved light extraction from emitters in high refractive index materials using solid immersion lenses. *J Appl Phys* 2002; **92**(2)**:** 660-665.

17. Frieder M, Jean-Christophe B. Electrowetting: from basics to applications. *Journal of Physics: Condensed Matter* 2005; **17**(28)**:** R705.

18. Lee CC, Hsiao SY, Fang W. Microlens formation technology utilizing multi-phase liquid ambiance. TRANSDUCERS 2009 - 2009 International Solid-State Sensors, Actuators and Microsystems Conference; 2009 21-25 June 2009; 2009. p. 2086-2089.

19. Jin X, Guerrero D, Klukas R, Holzman JF. Microlenses with tuned focal characteristics for optical wireless imaging. *Appl Phys Lett* 2014; **105**(3)**:** 031102.

20. Qian W, Ghislain LP, Elings VB. Imaging with solid immersion lenses, spatial resolution, and applications. *Proc IEEE* 2000; **88**(9)**:** 1491-1498.

21. Novoselov KS, Jiang D, Schedin F, Booth TJ, Khotkevich VV, Morozov SV, *et al.* Two-dimensional atomic crystals. *Proc Natl Acad Sci USA* 2005 July 26, 2005; **102**(30)**:** 10451-10453.

22. Mak KF, Lee C, Hone J, Shan J, Heinz TF. Atomically Thin $MoS_2$: A New Direct-Gap Semiconductor. *Phys Rev Lett* 2010 09/24/; **105**(13)**:** 136805.

23. RadisavljevicB, RadenovicA, BrivioJ, GiacomettiV, KisA. Single-layer MoS2 transistors. *Nat Nano* 2011 03//print; **6**(3)**:** 147-150.

24. Serrels KA, Ramsay E, Dalgarno PA, Gerardot B, O'Connor J, Hadfield RH, *et al.* Solid immersion lens applications for nanophotonic devices. *NANOP* 2008; **2**(1)**:** 021854-021854-021829.

25. Mouri S, Miyauchi Y, Matsuda K. Tunable Photoluminescence of Monolayer MoS2 via Chemical Doping. *Nano Letters* 2013 2013/12/11; **13**(12)**:** 5944-5948.

26. Li J, Wierzbowski J, Ceylan Ö, Klein J, Nisic F, Anh TL*, et al.* Tuning the optical emission of MoS2 nanosheets using proximal photoswitchable azobenzene molecules. *Appl Phys Lett* 2014; **105**(24)**:** 241116.

27. Buscema M, Steele GA, van der Zant HSJ, Castellanos-Gomez A. The effect of the substrate on the Raman and photoluminescence emission of single-layer MoS2. *Nano Research* 2014; **7**(4)**:** 561-571.

28. Scalise E, Houssa M, Pourtois G, Afanas'ev V, Stesmans A. Strain-induced semiconductor to metal transition in the two-dimensional honeycomb structure of MoS2. *Nano Research* 2012; **5**(1)**:** 43-48.

29. Amin B, Kaloni TP, Schwingenschlogl U. Strain engineering of WS2, WSe2, and WTe2. *RSC Advances* 2014; **4**(65)**:** 34561-34565.

30. Gao J, Li B, Tan J, Chow P, Lu T-M, Koratkar N. Aging of Transition Metal Dichalcogenide Monolayers. *ACS Nano* 2016 2016/02/23; **10**(2)**:** 2628-2635.